*Title*

A Staged Approach using Machine Learning and Uncertainty Quantification to Predict the Risk of Hip Fracture

*Authors*

Anjum Shaik[1#], Kristoffer Larsen[2#], Nancy E. Lane[3], Chen Zhao[4], Kuan-Jui Su[5], Joyce H. Keyak[6], Qing Tian[5], Qiuying Sha[2], Hui Shen[5], Hong-Wen Deng[5], Weihua Zhou[1,7]*

*Institutions*

[1.] Department of Applied Computing, Michigan Technological University, 1400 Townsend Dr, Houghton, MI, 49931

[2.] Department of Mathematical Sciences, Michigan Technological University, Houghton, MI, USA

[3.] Department of Internal Medicine and Division of Rheumatology, UC Davis Health, Sacramento, CA 95817

[4.] Department of Computer Science, Kennesaw State University, 680 Arntson Dr, Marietta, GA 30060

[5.] Division of Biomedical Informatics and Genomics, Tulane Center of Biomedical Informatics and Genomics, Deming Department of Medicine, Tulane University, New Orleans, LA 70112

[6.] Department of Radiological Sciences, Department of Biomedical Engineering, and Department of Mechanical and Aerospace Engineering, University of California, Irvine, CA, USA

[7.] Center for Biocomputing and Digital Health, Institute of Computing and Cybersystems, and Health Research Institute, Michigan Technological University, Houghton, MI 49931

# Anjum Shaik and Kristoffer Larsen contribute equally.

*\* Corresponding authors:*

Weihua Zhou, Ph.D.

Department of Applied Computing, Michigan Technological University,

1400 Townsend Dr, Houghton, MI, 49931, USA

Tel: 906-487-2666

E-Mail: whzhou@mtu.edu

# ABSTRACT

Hip fractures present a significant healthcare challenge, especially within aging populations, where they are often caused by falls. These fractures lead to substantial morbidity and mortality, emphasizing the need for timely surgical intervention. Despite advancements in medical care, hip fractures impose a significant burden on individuals and healthcare systems. This paper focuses on the prediction of hip fracture risk in older and middle-aged adults, where falls and compromised bone quality are predominant factors.

We propose a novel staged model that combines advanced imaging and clinical data to improve predictive performance. By using convolutional neural networks (CNNs) to extract features from hip DXA images, along with clinical variables, shape measurements, and texture features, our method provides a comprehensive framework for assessing fracture risk.

The study cohort included 547 patients, with 94 experiencing hip fracture. A staged machine learning-based model was developed using two ensemble models: Ensemble 1 (clinical variables only) and Ensemble 2 (clinical variables and DXA imaging features). This staged approach used uncertainty quantification from Ensemble 1 to decide if DXA features are necessary for further prediction. Ensemble 2 exhibited the highest performance, achieving an Area Under the Curve (AUC) of 0.9541, an accuracy of 0.9195, a sensitivity of 0.8078, and a specificity of 0.9427. The staged model also performed well, with an AUC of 0.8486, an accuracy of 0.8611, a sensitivity of 0.5578, and a specificity of 0.9249, outperforming Ensemble 1, which had an AUC of 0.5549, an accuracy of 0.7239, a sensitivity of 0.1956, and a specificity of 0.8343. Furthermore, the staged model suggested that 54.49% of patients did not require DXA scanning. It effectively balanced accuracy and specificity, offering a robust solution when DXA data acquisition is not always feasible. Statistical tests confirmed significant differences between the models, highlighting the advantages of the advanced modeling strategies.

Our staged approach offers a cost-effective holistic view of patients' health. It could identify individuals at risk with a high accuracy but reduce the unnecessary DXA scanning. Our approach has great promise to guide interventions to prevent hip fractures with reduced cost and radiation.

**Keywords**: Hip fracture, dual-energy X-ray absorptiometry, bone mineral density, machine learning, uncertainty quantification

**ABBREVATIONS**

| | |
|---|---|
| DXA | Dual-energy X-ray Absorptiometry |
| BMC | Bone Mineral Content |
| BMD | Bone Mineral Density |
| ML | Machine Learning |
| CNN | Convolutional Neural Network |
| AUC | Area Under the Receiver Operating Characteristic Curve |
| CI | Confidence Interval |
| RFE | Recursive Feature Elimination |

# INTRODUCTION

Hip fractures present a significant healthcare challenge, particularly among aging populations where they are often precipitated by falls. With the global aging trend, the incidence of hip fractures is expected to rise dramatically in the coming decades. For instance, while the annual global incidence was 1.3 million in 1990, it is projected to surge to a staggering 7 to 21 million by 2050 [1]. In the United States alone, the annual incidence per 100,000 individuals ranges between 197 to 201 for men and 511 to 553 for women, with rates increasing significantly with age [2]. These incidents have serious consequences on quality of life. Apart from causing morbidity and mortality, hip fractures impose a substantial economic burden. Patients often face approximately $40,000 in expenses within the first-year post-fracture, while the collective annual cost in the US alone surpasses $17 billion.

Diagnosing hip fractures hinges on a meticulous clinical evaluation, typically initiated by history of falls resulting in hip pain and restricted mobility. However, the diagnostic process extends beyond mere fracture identification, encompassing a comprehensive assessment of underlying medical conditions, social circumstances, and cognitive function, all of which profoundly impact patient care and prognosis. While surgical intervention remains the primary treatment modality for most hip fractures, the timing of surgery is critical for effective pain management and functional restoration. Ensuring optimal preparation for surgery, especially among elderly individuals with intricate medical requirements, is essential for minimizing perioperative risks and maximizing postoperative outcomes. Thus, the diagnostic and treatment approach for hip fractures encompasses not only fracture management but also comprehensive patient-centered care aimed at maximizing functional recovery and quality of life [3].

Bone mineral density (BMD) is a key determinant of hip fracture risk. Dual-energy X-ray absorptiometry (DXA) plays a pivotal role in assessing BMD and fracture risk. DXA serves as the standard imaging modality guiding clinical decisions for the detection, initiation of treatment, and follow-up of individuals with osteoporosis and fracture risk. Recent studies have explored innovative approaches, such as artificial intelligence (AI), to enhance hip fracture risk prediction by leveraging DXA imaging alongside clinical data. Lex et al.[4] conducted a thorough investigation into the diagnostic accuracy of artificial intelligence (AI) models in diagnosing hip fractures on radiographs and predicting postoperative clinical outcomes following hip fracture surgery relative to current practices. Their systematic review and meta-analysis of 39 studies revealed that AI models perform comparably to expert clinicians in diagnosing hip fractures. Cha et al.[5] systematically reviewed the use of AI and machine learning (ML) in diagnosing and classifying hip fractures, demonstrating high accuracy and effectiveness in clinical settings. Furthermore, Murphy et al.[6] utilized two sets of radiographs: one from population without hip fractures collected as part of a bone mass study and another from those who had hip fractures from local National Hip Fracture Database (NHFD) audit records. Their study demonstrated that a trained neural network exhibits a remarkable 19% increase in accuracy in classifying hip fractures compared to experienced human observers within clinical settings. This finding underscores the transformative impact of ML technologies in augmenting the capabilities of healthcare professionals and improving patient outcomes in orthopedic care. Zhao et al.[7] introduced multi-view variational autoencoder (MVAE) and product of expert (PoE) models for predicting proximal femoral fracture loads by integrating whole-genome sequence features and DXA-derived imaging features. Additionally, Hong et al.[8] developed a bone radiomics score using a random forest model and texture analysis of DXA hip images, in predicting incident hip fractures. Despite advancements, current ML and AI approaches for predicting hip fractures have notable limitations. Some often utilize only a single modality of data, either clinical or imaging, which can lead to limited predictive accuracy. Moreover, most multi-modality ML methods require all modalities to be obtained in advance for effective prediction, adding the cost, radiation and complexity to the diagnostic process. Notably, the actual process of obtaining these modalities perhaps can be made sequential in clinic practice depending on our results in order to be more efficient and economical.

To address these limitations, our study introduces a novel staged modeling approach aimed at predicting hip fractures. Unlike current methods, our approach is structured into two distinct stages. In the first stage, we focus solely on clinical features. Subsequently, in the second stage, we expand our analysis to incorporate imaging features extracted from hip DXA images. By integrating both clinical and imaging data with ML and uncertainty quantification, this staged approach aims to enhance prediction accuracy and adaptability to diverse clinical scenarios.

## MATERIALS AND METHODS

The dataset utilized in this study was sourced from the UK Biobank (application ID: 61915), representing a valuable resource for investigating bone health parameters. DXA imaging, essential for evaluating BMD and morphology, was performed by trained radiographers using the GE-Lunar iDXA instrument. Regular calibration of this instrument to a manufacturer's phantom (GE-Lunar, Madison, WI) and daily quality control procedures ensured the accuracy and reliability of DXA measurements [9]. This comprehensive DXA dataset covers various anatomical regions, including the whole body, lateral thoraco-lumbar spine, and bilateral hips and knees. Focusing on a subset of 547 patients with DXA hip images, we manually annotated the left and right contours of femur. to isolate the femur, the region of interest, from the raw images. Additionally, we incorporated other relevant clinical features into our analysis. Among the subset of patients with DXA hip images, 94 individuals had fractures (including 40 males), while the majority (n= 453) were non-fractured individuals (with 226 males). The ethnicity of all patients in our sample is British.

### Clinical factors

Our study considered a plethora of variables crucial for understanding various aspects of participants' health profiles. These variables encompassed demographic details such as age at recruitment, sex, and genetic sex, alongside anthropometric measurements like weight. Additionally, information regarding participants' average total household income before tax, and lifestyle factors including smoking and alcohol consumption statuses were considered (Supp. Table 1). We also examined dietary habits, including the variation in diet and major dietary changes in the last 5 years, along with occurrences of falls and bone fractures in the past year and 5 years, respectively. Furthermore, the dataset also provided intricate measurements of BMD and bone mineral content (BMC) at various anatomical sites, shedding light on participants' bone health. The regular intake of vitamin and mineral supplements was also documented. This comprehensive array of clinical data facilitated a thorough exploration of factors influencing participants' health and enabled meaningful insights into bone health and related risk factors.

**Model evaluation**

In our study focused on predicting hip fracture risk, we propose an advanced staged modeling approach meticulously designed to enhance predictive accuracy while concurrently minimizing clinical costs and procedural time. Inspired by the sequential decision-making processes commonly observed in clinical practice, our methodology incorporates advanced techniques to optimize model performance (Figure 1.). To ensure consistency across the dataset, DXA images are standardized to a size of 224 pixels. Our model evaluation process begins with an analysis of hip DXA images, which are meticulously annotated to delineate anatomical outlines, serving as the primary input data. Additionally, we integrate crucial clinical data, including age, weight, sex, alcohol consumption, smoking status, dietary changes due to illness in the last 5 years, how often diet varies week to week, falls within the last year, history of fractured or broken bones in the last 5 years, and average household income, into our analysis.

Feature extraction is performed using two pre-trained CNN models: VGG16 [10] and Xception [11]. These CNN models are adept at extracting rich feature representations from the preprocessed DXA images, capturing both global and fine-grained details crucial for accurate prediction. Alongside CNN-based feature extraction, 2D shape measurements and texture features from the DXA images using specialized packages were computed. Specifically, the shape measurements are computed using the IMEA package [12], which assess the 2D geometric characteristics of the femur region. Similarly, texture features are extracted using the PyRadiomics package [13], enabling the capture of detailed textural information from the DXA images. Subsequently, the extracted features from the CNN models, shape measurements, texture features and clinical data are combined to form a comprehensive feature set for model evaluation. This integrated feature set provides a holistic representation of both anatomical and clinical aspects relevant to hip fracture prediction.

To mitigate dimensionality and enhance model interpretability, feature selection techniques such as univariate feature selection via near zero variance filtering and correlation filtering are employed. Furthermore, Recursive Feature Elimination (RFE) [14] was used to identify the most relevant features in a multivariate fashion. These methods identify the most relevant features for predicting the target variable, ensuring that only informative features are retained for analysis. Bootstrapping is utilized to create diverse ensembles of models for each stage of the sequential modeling process. Each ensemble model is trained on a resampled subset of the data, promoting robustness, and capturing variations within the dataset. A sequential model is then constructed to integrate predictions from different stages, leveraging the strengths of multiple collections of sub-models.

In the evaluation of our model, a nested cross-validation structure was employed (Figure 2.). Initially, the dataset was divided into an outer training fold comprising 492 samples and an outer test fold with 55 samples. From the training fold, two separate validation sets were extracted, each containing 45 samples. To preprocess the data and mitigate outliers, centering/scaling and spatial sign techniques were applied. Next, the two validation sets were sliced from the original training fold to fine-tune the hyperparameters of the staged model. These hyperparameters, including standard deviation threshold and midway thresholds, govern the transition from stage 1 to stage 2 in the model, i.e., whether a patient will need to acquire DXA images for a new prediction.

Inner cross-validation was conducted on the remaining training data to optimize the hyperparameters of ensemble models, such as the number of base models comprising the ensembles, the percentage of random samples for training each base model, and the respective base models hyperparameters. This process was repeated twice, once for stage 1 data (clinical features) and once for stage 2 data (clinical and DXA image features), resulting in two ensemble models: ensemble 1 and ensemble 2.

After training the ensemble models separately, the first validation set was utilized to fine-tune the hyperparameter thresholds of the staged model. These thresholds were optimized to achieve a balanced trade-off between predictions retained from Ensemble 1 and those cascading into Ensemble 2, using a scaled weighted Area Under the Curve (AUC) metric. Finally, the best-performing staged model was evaluated using the outer test set to ensure its robustness and generalization.

**STATISTICAL ANALYSIS**

In this study, comprehensive statistical analyses were performed to evaluate model performance and feature associations with hip fracture risk. The DeLong test was used to compare ROC curves and McNemar's test assessed sensitivity and specificity variations. Chi-square test and Fisher's exact tests revealed the associations between categorical variables and fracture risk, while t-tests highlighted differences in continuous variables between fracture groups.

**RESULTS**

The study cohort comprised 547 patients, with 94 individuals having previously experienced hip fractures. An initial assessment revealed that 54.49% of the patients did not require DXA scanning, while 45.52% did. Patients were classified as not requiring DXA scanning based on the absence of significant risk factors such as younger age, no history of fractures, absence of clinical risk factors for osteoporosis (e.g., history of smoking, excessive alcohol consumption), and initial clinical assessments indicating low risk. The distribution of patients not requiring DXA was characterized by the following percentiles: 25th percentile at 35.45%, 50th percentile at 46.78%, and 75th percentile at 59.55%.

Table 1 represents the performance metrics of the models employed in this study. Notably, Ensemble 2 emerged as the frontrunner with the highest AUC of 0.95 (95% CI: 0.87-1.00), followed closely by the staged model at 0.85 (95% CI: 0.78-0.92). Ensemble 1 exhibited a comparatively lower AUC of 0.70 (95% CI: 0.55-0.85). These findings underline the superior predictive performance of Ensemble 2 and the staged model in fracture risk assessment. Diving deeper into accuracy and specificity, the staged model showcased superior performance, with accuracy reaching 86.11% and specificity peaking at 92.49%. Although Ensemble 2 displayed commendable sensitivity (80.78%), its accuracy and specificity were outmatched by the staged model. Additionally, fracture risk assessment tool (FRAX) with BMD and FRAX without BMD yielded AUC scores of 0.7577 and 0.6185, respectively. Importantly, all models showcased marked improvements over the guideline model, signifying the efficacy of advanced ML approaches in fracture risk assessment.

Furthermore, the analysis involved rigorous statistical testing to compare the performance of various models. DeLong tests and McNemar's sensitivity and specificity test were utilized, revealing significant differences between the models. Confidence intervals for the DeLong tests were computed, indicating the range of AUC values with 95% confidence. For example, the staged model had a 95% CI of 0.8083-0.8893, Ensemble 1 had a 95% CI of 0.4882-0.6135, Ensemble 2 had a 95% CI of 0.9388-0.976, FRAX with BMD had a 95% CI of 0.7002-0.8151, and FRAX without BMD had a 95% CI of 0.551-0.686. Additionally, the DeLong tests yielded p-values, indicating the significance of the differences in AUC between different model pairs. For instance, the p-value for comparing Staged vs. Ensemble 1 was <0.001, and for comparing Staged vs. FRAX with BMD it was 0.0041. McNemar's sensitivity and specificity test also provided insights, with p-values indicating the significance of differences in sensitivity and specificity between model pairs, such as between Staged and Boot1 (sensitivity: <0.0001, specificity: <0.0001).

Additionally, the study identified significant associations between categorical variables (Table 2) like alcohol consumption and average household income with fracture risk, as well as notable differences in continuous variables such as age and various BMD measurements among patient groups (Table 3). Baseline statistics including p-values from chi-square, Fisher , and ttests for categorical (Table 4) and continuous (Table 5) variables were also calculated. These findings underscore the potential of ensemble learning and staged modeling in enhancing hip-fracture risk assessment, offering insights for clinical decision-making and preventive strategies.

To visually encapsulate the findings, the AUC curves of the ensemble stage 1 (Figure 3A), ensemble stage 2 (Figure 3B) and staged (Figure 3C) models are presented. Ensemble 2 emerged as the standout performer, consistently surpassing its counterparts. However, no significant disparities were observed between the staged model and either Ensemble 1 or 2, underscoring the robustness of the staged approach. Figures 4A and 4B further enrich our understanding by highlighting the importance of various features. Ensemble models underscored age, weight, and dietary changes as significant predictors (Figure 4A). Conversely, Ensemble 2 prioritized DXA parameters, such as convex area and projection area, accentuating their role in fracture risk assessment (Figure 5).

# DISCUSSION

In our study, we developed a staged based ML model to predict hip fractures, utilizing data obtained from 547 patients, including 94 individuals with a history of hip fractures from the UK Biobank dataset. Ensemble model 1 included only clinical features while ensemble model 2 included DXA image feature along with clinical features. The staged model demonstrated comparable performance to Ensemble 2, which incorporated both clinical and DXA features, with an AUC of 0.85 compared to 0.95, accuracy of 0.86 compared to 0.92, sensitivity of 0.67 compared to 0.80, and specificity of 0.92 compared to 0.94, respectively, however, the staged model only utilized DXA data 45.52% of the time.

**AI for hip fracture risk prediction**

Recent advancements in hip fracture risk prediction have been marked by a notable transition towards the incorporation of AI and machine learning ML techniques. Researchers such as Twinprai et al.[15] focused on the diagnostic accuracy of a YOLOv4-tiny AI model for classifying hip fractures from radiographic images. Their model achieved a sensitivity of 96.2%, specificity of 94.6%, and accuracy of 95%, significantly outperforming general practitioners and first-year residents, and matching the performance of specialist doctors. This demonstrates the potential of AI in enhancing diagnostic precision and efficiency. Li et al. [16] developed a risk prediction model using a Random Survival Forest (RSF) algorithm to predict long-term mortality post-hip fracture surgery, achieving a C statistic of 0.83 for 30-day and 0.75 for 1-year mortality. Their model identified key risk factors such as post-operative complications, age, and pre-existing conditions, providing a robust framework for predicting patient outcomes over extended periods. Xu et al.[17] utilized three ML models (Random Forest, Extreme Gradient Boosting, and Backpropagation Neural Network) to predict in-hospital mortality in patients with severe femoral neck fractures, achieving AUC values of 0.98, 0.97, and 0.95, respectively. These high AUC values demonstrate the efficacy of ML models in predicting critical outcomes and guiding early clinical decision-making. The integration of AI and ML technologies in hip fracture diagnosis and mortality prediction signifies a significant stride forward in orthopedic care. These advanced models offer enhanced precision and efficiency in clinical decision-making, enabling early detection and personalized treatment strategies for hip fracture patients. Despite these advancements, challenges persist, particularly in integrating multi-modal data and interpreting complex AI-driven models. One significant challenge lies in the reliance on large and diverse datasets for training and validation, which may not always be readily accessible in clinical settings. Moreover, the interpretability of AI-driven models remains a concern, as their complex algorithms often lack transparency, hindering clinicians' understanding of prediction rationales. Additionally, while ML models demonstrate high accuracy and effectiveness in controlled research environments, their real-world applicability and generalizability to diverse patient populations necessitate further exploration and validation.

**Staged modeling for hip fracture risk prediction**

Our staged approach for hip fracture risk prediction represents a novel methodology aimed at enhancing the accuracy and reliability of fracture risk assessment. Unlike traditional single-stage models, which often rely on a singular set of features for prediction, our approach systematically integrates multiple stages, each tailored to leverage specific types of data. In the first stage of our staged approach, we focus on utilizing clinical variables to build a foundational understanding of each patient's health profile. This initial stage incorporates demographic details, medical history, lifestyle factors, and other relevant clinical indicators to establish a comprehensive baseline for fracture risk assessment. Following the initial clinical assessment, our approach progresses to the ensemble stage 2, where imaging features extracted from hip DXA images are included. By incorporating this additional layer of data, we aim to enrich the predictive capabilities of our model, capturing subtle nuances and anatomical insights that may not be discernible from clinical variables alone. Ensemble 2 emerged as the top-performing model, achieving a high AUC of 0.95. The model also had an average accuracy of 0.9195. Its sensitivity (0.8078) and specificity (0.9427) were notably high. In assessing the performance of the stage 2 model within our staged framework, we scrutinized its AUC alongside corresponding confidence intervals relative to standard deviation percentiles (Figure 5.). As our analysis progressed from left to right along these percentiles which results in a smaller and smaller subset of the data whose patients have higher uncertainty, we notice the performance of the model in terms of AUC decreases. This approach allows us to delve into predictions with higher uncertainty, showcasing that increased uncertainty leads to decreased performance. Moreover, this opens up room to add a third stage potentially to include genetic data [18] or QCT images [19] which are more costly and less available than DXA but can provide more nuanced complementary information in the sequential approach.

One of the key strengths of our staged approach lies in its adaptability and flexibility. The use of internal logic rules allows for dynamic decision-making, determining whether the acquisition of DXA data is necessary based on the information gathered in the initial clinical stage. This ensures that resources are allocated efficiently, with additional imaging studies being performed only when deemed essential for further risk assessment. Moreover, our staged approach offers enhanced interpretability compared to complex AI-driven models. By breaking down the prediction process into distinct stages, clinicians can better understand the rationale behind each decision, facilitating trust and confidence in the model's outputs.

**Comparison with FRAX**

In comparing the performance of our staged approach for hip fracture risk prediction with the FRAX tool, we observe notable differences in predictive accuracy. Our approach, leveraging a combination of clinical variables and imaging features, achieved an AUC of 0.85, demonstrating superior discriminatory ability compared to FRAX. Specifically, when comparing our approach to FRAX with BMD, which attained an AUC of 0.7577, we find that our model outperformed it significantly. The higher AUC value of our approach indicates enhanced sensitivity and specificity in identifying individuals at risk of hip fractures, thereby improving the overall predictive performance. Similarly, when comparing our approach to FRAX without BMD, which yielded an AUC of 0.6185, our model again exhibited superior performance. Despite FRAX being a widely used tool for fracture risk assessment, our staged approach demonstrated enhanced accuracy and reliability in predicting hip fractures, underscoring the effectiveness of incorporating imaging features alongside clinical variables.

**Cost and radiation reduction**

The staged approach for hip fracture risk prediction offers a comprehensive strategy that not only enhances diagnostic accuracy but also addresses cost and radiation concerns associated with conventional methods. It revealed that 54.49% of the patients did not require DXA scanning, while 45.52% did. The distribution of patients not requiring DXA was characterized by the following percentiles: 25th percentile at 35.45%, 50th percentile at 46.78%, and 75th percentile at 59.55%. By accurately identifying individuals at high risk of hip fractures, our approach enables targeted intervention and preventive measures, minimizing unnecessary diagnostic tests and treatments for those at lower risk. This tailored approach optimizes resource allocation, leading to significant cost savings within healthcare systems. Furthermore, early detection and intervention facilitated by our approach can prevent costly hip fracture-related complications, such as prolonged hospital stays and postoperative issues, thus reducing overall healthcare expenditure. An essential aspect of our approach is the incorporation of imaging features extracted from existing diagnostic scans, such as DXA images. This eliminates the need for additional imaging tests, thereby minimizing radiation exposure for patients. Leveraging existing imaging data more efficiently not only prioritizes patient safety but also mitigates potential risks associated with excessive radiation exposure, including long-term health consequences. Moreover, our staged modeling approach allows for the selective use of advanced imaging techniques, like DXA scans, based on individual risk profiles derived from clinical data. This targeted approach minimizes the need for unnecessary imaging tests, further reducing radiation exposure and associated costs. Additionally, the integration of clinical as well as DXA imaging data provides a holistic assessment of fracture risk, enhancing diagnostic accuracy and reducing the likelihood of missed diagnoses or unnecessary treatments.

**Interpretability of our model**

The clinical relevance of our findings is underscored by the identification of significant predictors of hip fracture risk. Our models identified age, weight, dietary changes, and DXA parameters as key predictors, aligning with established literature on fracture risk factors. These findings have the potential to guide clinical decision-making by enabling the early identification of individuals at high risk of fractures, thus facilitating the implementation of tailored interventions to effectively reduce fracture risk. There is potential for further refinement and expansion of our staged modeling approach by incorporating additional features, such as genetic data, to enhance the predictive capabilities of the model.

## LIMITATIONS

First, this study involved a relatively small number of subjects, which is an inherent limitation in the study design. Increasing the sample size would improve the statistical power and generalizability of our findings. Additionally, the performance and feasibility of the data-driven system might be influenced by the quality of the data. For instance, inconsistencies or inaccuracies in clinical data and DXA images could impact the model's predictive accuracy. Second, there were missing features in the UKBiobank repository. Not all potential risk factors for hip fractures were captured or included in the analysis. This limitation might have resulted in an incomplete representation of each patient's health profile. Incorporating additional relevant features such as genetic data, comprehensive environmental factors, and more detailed medical history could further refine the model's predictive capabilities. Lastly, this study did not include external validation using datasets from other populations or healthcare settings.

## CONCLUSION

We developed a staged approach combining clinical data and DXA hip images for hip fracture risk prediction. By considering various factors like age, weight, and bone health alongside images with machine learning and uncertainty quantification, the model offers a cost-effective holistic view of patients' health. Through rigorous evaluation, we found that our staged approach could identify individuals at risk with a high accuracy but reduce the unnecessary DXA scanning. It has great promise to guide interventions to prevent hip fractures with reduced cost and radiation.

**Acknowledgments**

This research has been conducted using the UK Biobank Resource under application number [61915]. It was supported in part by grants from the National Institutes of Health, USA (P20GM109036, R01AR069055, U19AG055373, R01AG061917, and R15HL172198) and NASA Johnson Space Center, USA contracts NNJ12HC91P and NNJ15HP23P.

*Conflict of Interest Disclosure Statement*

All authors declare that there are no conflicts of interest.

# REFERENCES

[1] Gullberg B, Johnell O, Kanis JA. World-wide Projections for Hip Fracture. Osteoporos Int. 1997 Sep 1;7(5):407–413.

[2] Dhanwal DK, Dennison EM, Harvey NC, Cooper C. Epidemiology of hip fracture: Worldwide geographic variation. Indian J Orthop. 2011 Jan;45(1):15–22. PMCID: PMC3004072

[3] Emmerson BR, Varacallo M, Inman D. Hip Fracture Overview. StatPearls [Internet]. Treasure Island (FL): StatPearls Publishing; 2024 [cited 2024 May 20]. Available from: http://www.ncbi.nlm.nih.gov/books/NBK557514/ PMID: 32491446

[4] Lex JR, Di Michele J, Koucheki R, Pincus D, Whyne C, Ravi B. Artificial Intelligence for Hip Fracture Detection and Outcome Prediction. JAMA Netw Open. 2023 Mar 17;6(3):e233391. PMCID: PMC10024206

[5] Cha Y, Kim J-T, Park C-H, Kim J-W, Lee SY, Yoo J-I. Artificial intelligence and machine learning on diagnosis and classification of hip fracture: systematic review. J Orthop Surg Res. 2022 Dec 1;17(1):520. PMCID: PMC9714164

[6] Murphy EA, Ehrhardt B, Gregson CL, von Arx OA, Hartley A, Whitehouse MR, Thomas MS, Stenhouse G, Chesser TJS, Budd CJ, Gill HS. Machine learning outperforms clinical experts in classification of hip fractures. Sci Rep. Nature Publishing Group; 2022 Feb 8;12(1):2058.

[7] Zhao C, Keyak JH, Cao X, Sha Q, Wu L, Luo Z, Zhao L, Tian Q, Qiu C, Su R, Shen H, Deng H-W, Zhou W. Multi-view information fusion using multi-view variational autoencoders to predict proximal femoral strength [Internet]. arXiv; 2023 [cited 2023 Oct 31]. Available from: http://arxiv.org/abs/2210.00674

[8] Hong N, Park H, Kim CO, Kim HC, Choi J-Y, Kim H, Rhee Y. Bone Radiomics Score Derived From DXA Hip Images Enhances Hip Fracture Prediction in Older Women. J Bone Miner Res. 2021 Sep;36(9):1708–1716. PMID: 34029404

[9]: Resource 502 [Internet]. [cited 2024 May 20]. Available from: https://biobank.ctsu.ox.ac.uk/crystal/refer.cgi?id=502

[10] Simonyan K, Zisserman A. Very Deep Convolutional Networks for Large-Scale Image Recognition [Internet]. arXiv.org. 2014 [cited 2024 May 20]. Available from: https://arxiv.org/abs/1409.1556v6

[11] Chollet F. Xception: Deep Learning with Depthwise Separable Convolutions [Internet]. arXiv; 2017 [cited 2024 May 20]. Available from: http://arxiv.org/abs/1610.02357

[12] Kroell N. imea: A Python package for extracting 2D and 3D shape measurements from images. Journal of Open Source Software. 2021 Apr 6;6(60):3091.

[13] van Griethuysen JJM, Fedorov A, Parmar C, Hosny A, Aucoin N, Narayan V, Beets-Tan RGH, Fillion-Robin J-C, Pieper S, Aerts HJWL. Computational Radiomics System to Decode the Radiographic Phenotype. Cancer Research. 2017 Oct 31;77(21):e104–e107.

[14] Guyon I, Weston J, Barnhill S, Vapnik V. Gene Selection for Cancer Classification Using Support Vector Machines. Machine Learning. 2002 Jan 1;46:389–422.

[15] Twinprai N, Boonrod A, Boonrod A, Chindaprasirt J, Sirithanaphol W, Chindaprasirt P, Twinprai P. Artificial intelligence (AI) vs. human in hip fracture detection. Heliyon. 2022 Oct 27;8(11):e11266. PMCID: PMC9634369

[16] Li Y, Chen M, Lv H, Yin P, Zhang L, Tang P. A novel machine-learning algorithm for predicting mortality risk after hip fracture surgery. Injury. 2021 Jun 1;52(6):1487–1493.

[17] Xu L, Liu J, Han C, Ai Z. The Application of Machine Learning in Predicting Mortality Risk in Patients With Severe Femoral Neck Fractures: Prediction Model Development Study. JMIR Bioinformatics and Biotechnology. 2022 Aug 19;3(1):e38226.

[18] Nethander M, Coward E, Reimann E, Grahnemo L, Gabrielsen ME, Wibom C, Estonian Biobank Research Team, Mägi R, Funck-Brentano T, Hoff M, Langhammer A, Pettersson-Kymmer U, Hveem K, Ohlsson C. Assessment of the genetic and clinical determinants of hip fracture risk: Genome-wide association and Mendelian randomization study. Cell Rep Med. 2022 Oct 18;3(10):100776. PMCID: PMC9589021

[19] Awal R, Faisal T. QCT-based 3D finite element modeling to assess patient-specific hip fracture risk and risk factors. Journal of the Mechanical Behavior of Biomedical Materials. 2024 Feb 1;150:106299.

# FIGURES AND TABLES

## Figure 1. Staged process

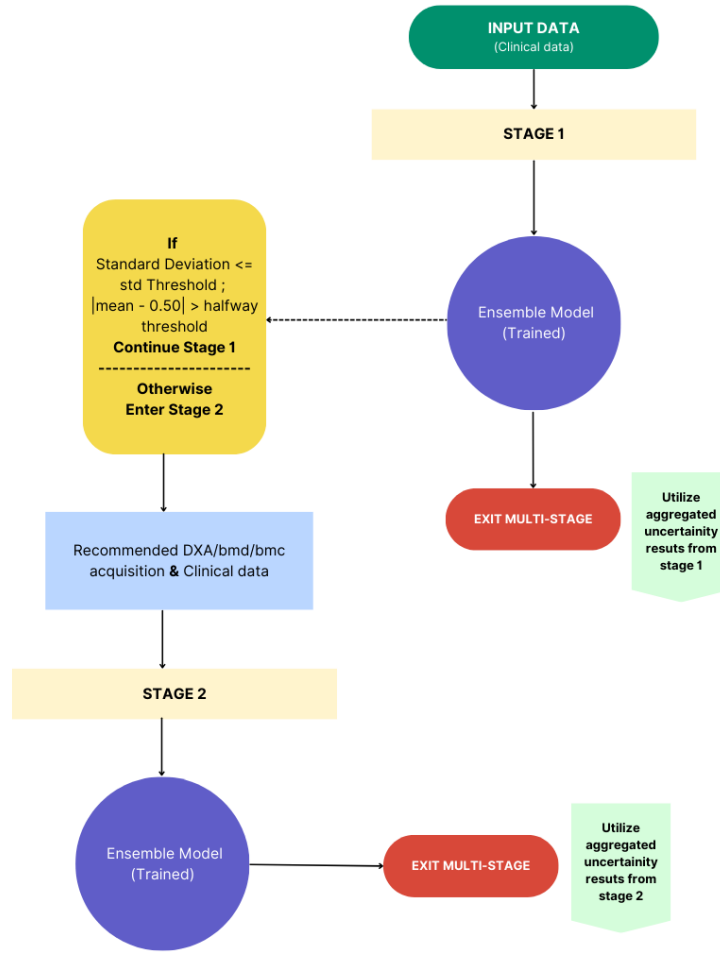

Figure 1 illustrates the staged approach for hip fracture prediction, employing internal logic rules where the model progresses to Stage 1 if the standard deviation is less than or equal to the specified threshold and the absolute difference between the mean and 0.50 exceeds the halfway threshold; otherwise, Stage 2 is initiated for comprehensive risk evaluation.

**Figure 2. Data modelling process**

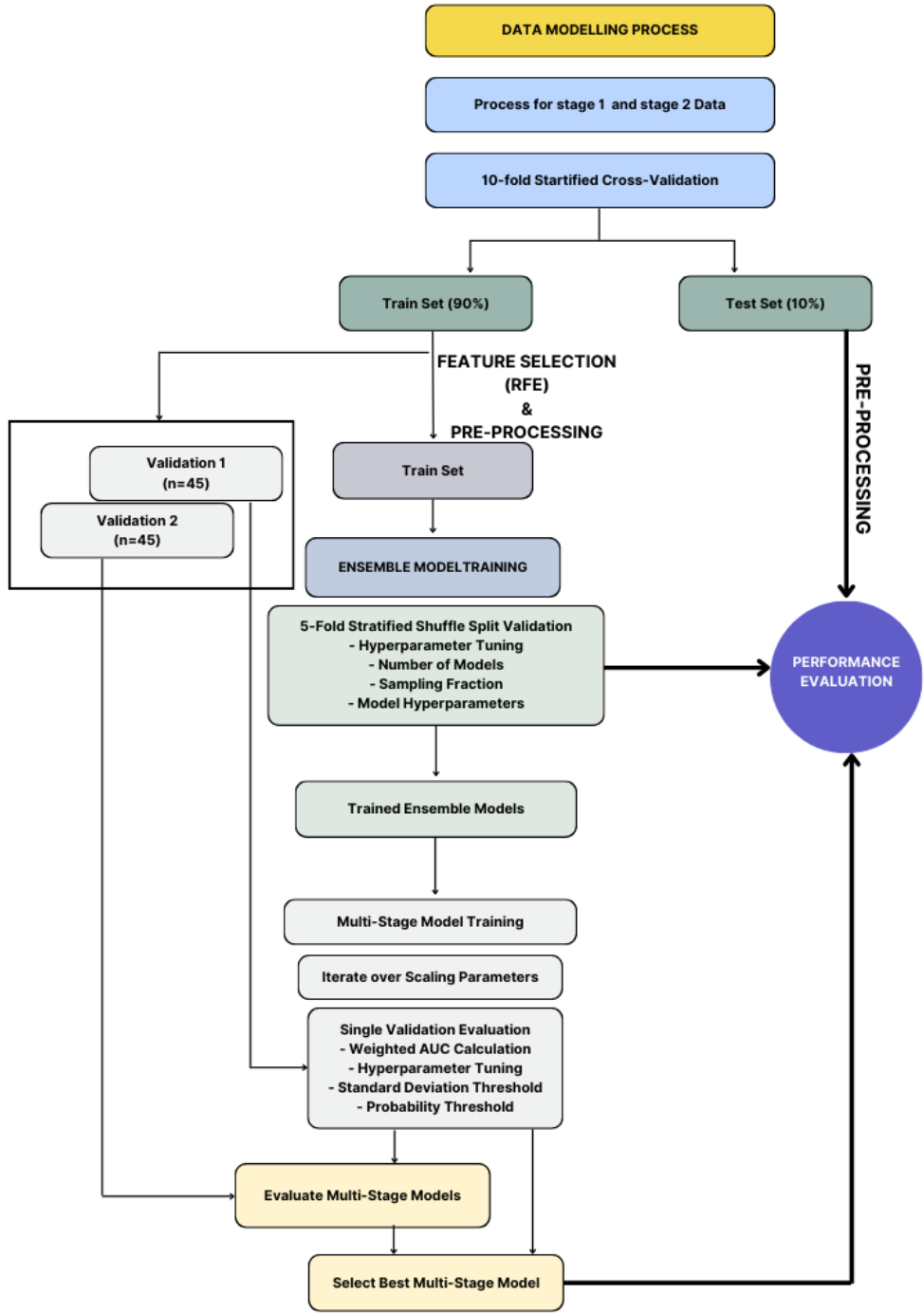

Figure 2 illustrates the modeling process, encompassing feature extraction, selection, and ensemble techniques, to optimize predictive performance using clinical and imaging data, followed by evaluation and validation of the resulting model.

**Figure 3A. ROC curves for ensemble stage 1**

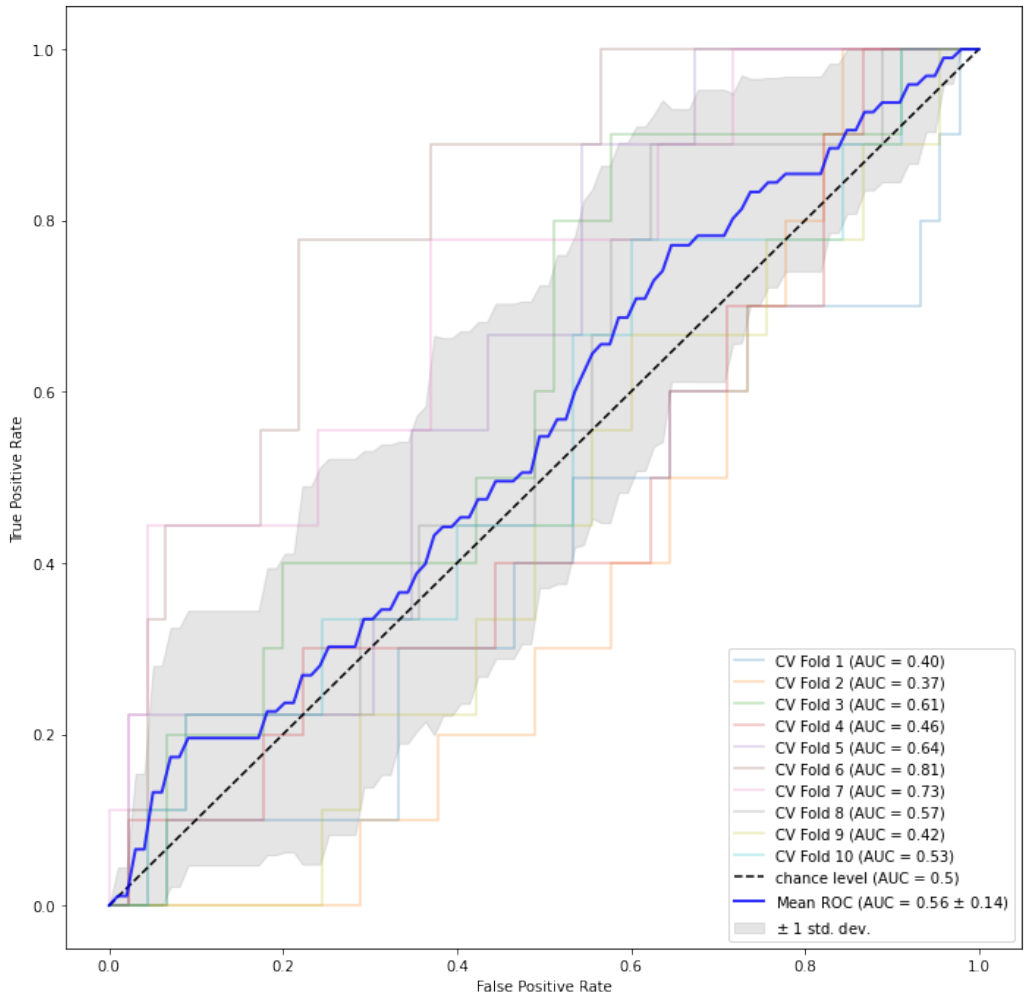

Figure 3A plot depicts the mean ROC curve representing the average performance of Ensemble Stage 1 in predicting hip fracture risk, along with variability represented by standard deviation. The ROC curve illustrates the trade-off between sensitivity and specificity across different threshold values.

**Figure 3B. ROC curves for ensemble stage 2**

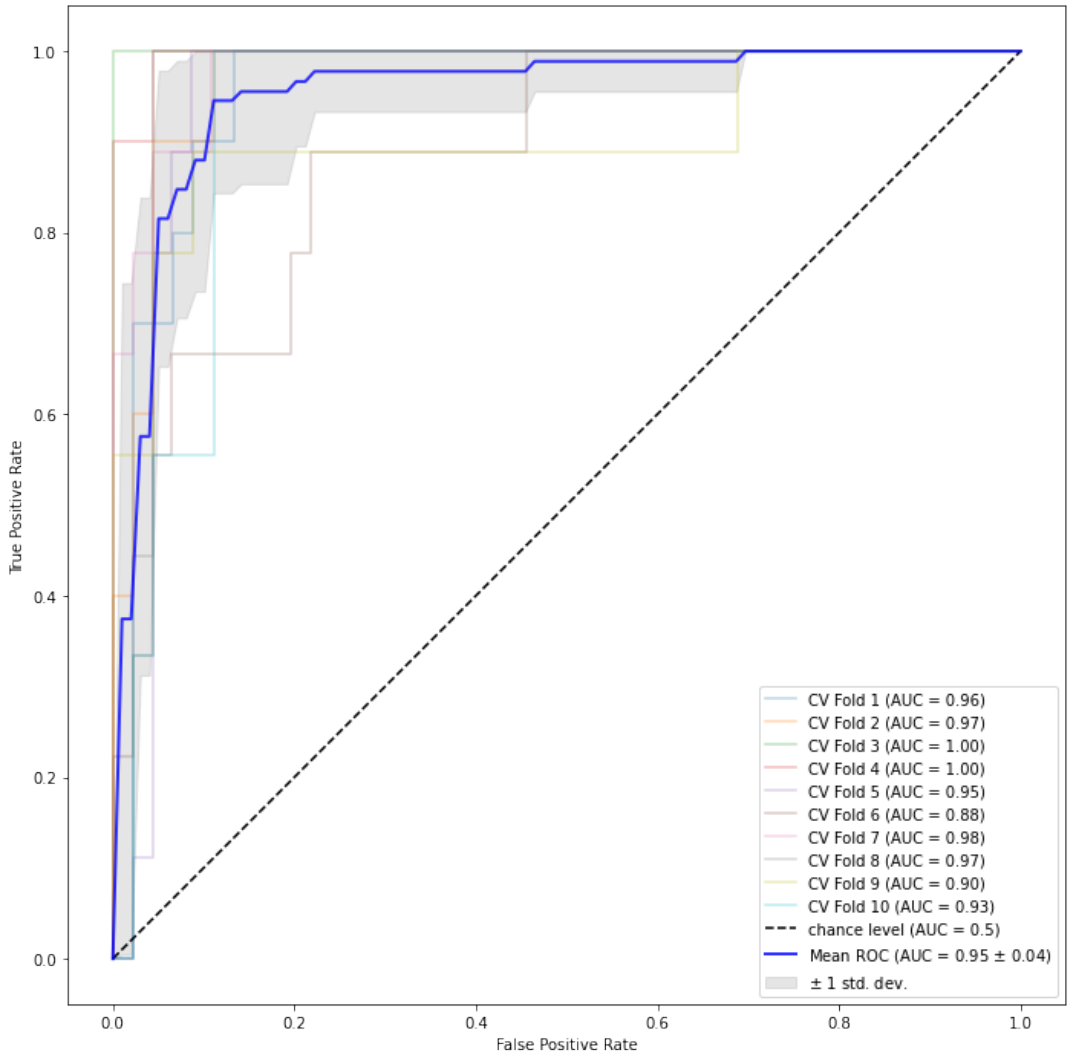

Figure 3B. The plot shows the mean ROC curve for Ensemble Stage 2, depicting the average performance across multiple iterations. The variability around the mean curve illustrates the uncertainty associated with the model's predictions.

**Figure 3C. ROC curves for the staged model**

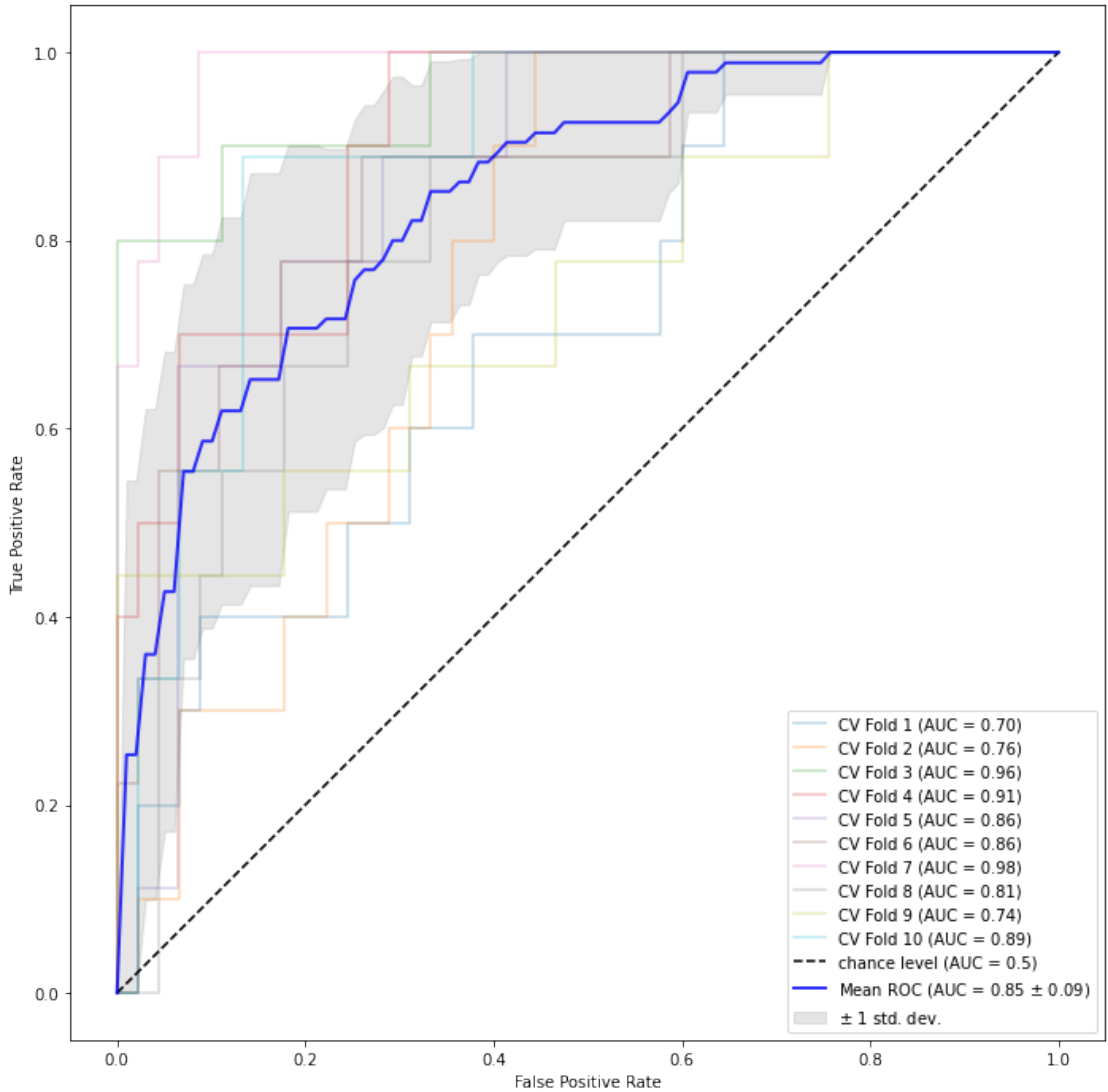

Figure 3C depicts Receiver Operating Characteristic (ROC) curves for the staged hip fracture risk prediction model. The mean ROC curve represents the average performance across multiple iterations, with the shaded area indicating the variability or uncertainty associated with the model's predictions.

## Figure 4A. Feature importance – Ensemble stage 1

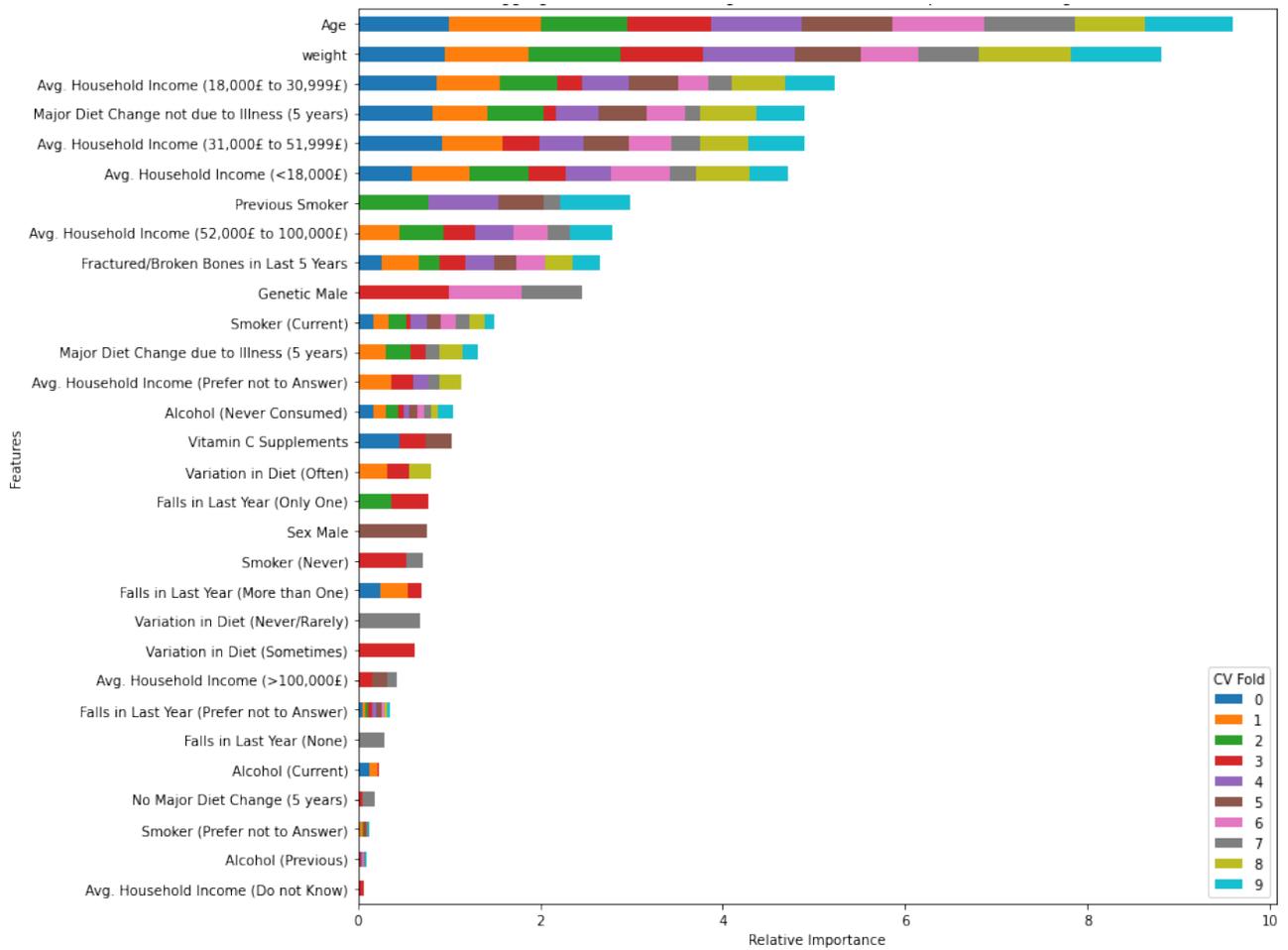

Figure 4A illustrates the fold-aggregated ensemble-averaged absolute feature importance for stage 1 ensemble models. This assessment determines the importance of each feature based on its contribution to the predictive performance of the ensemble model.

**Figure 4B. Feature importance – Ensemble stage 2**

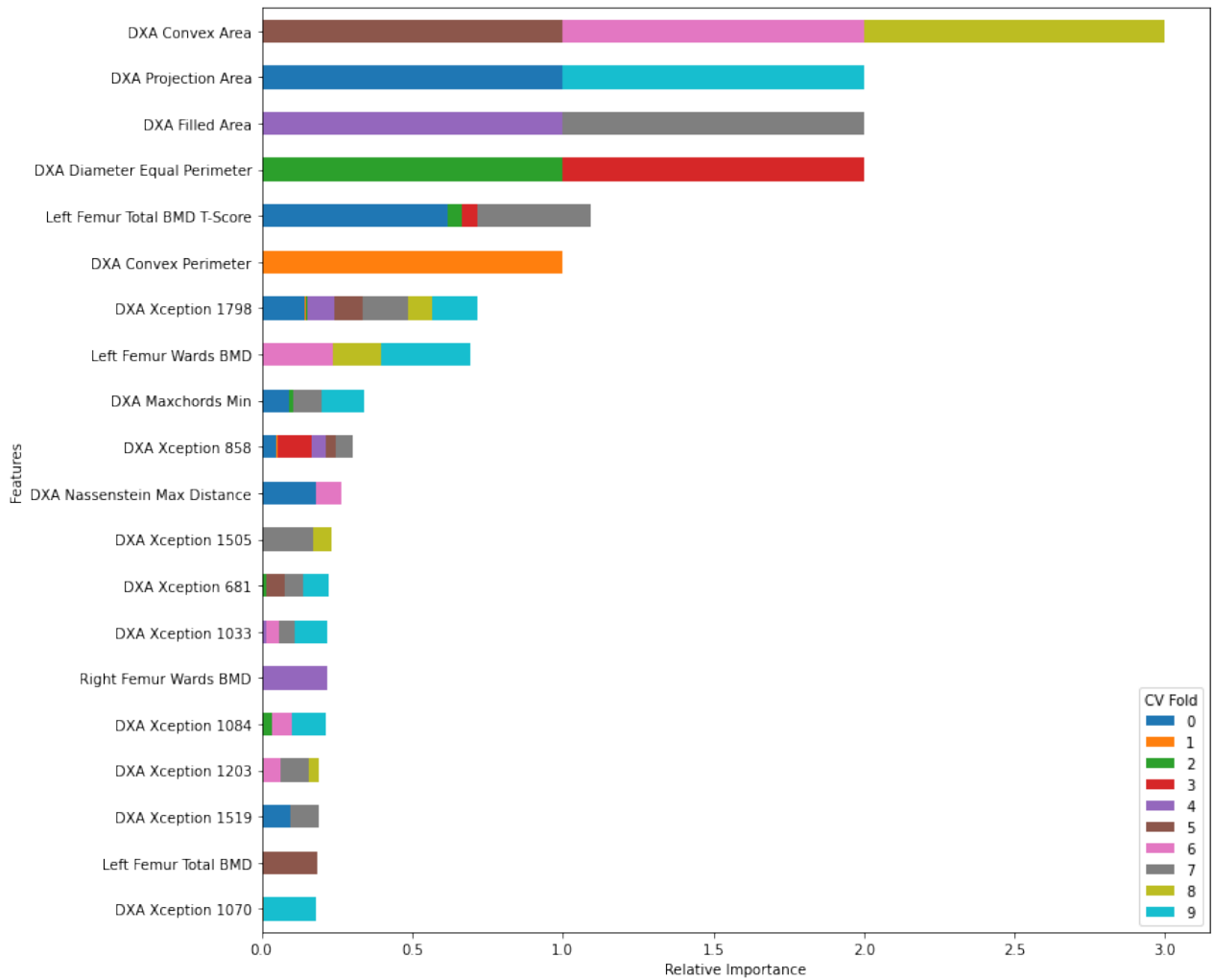

Figure 4B showcases the top 20-fold aggregated ensemble averaged absolute feature importance for Stage 2 ensemble modeling. The importance of each feature is determined based on its contribution to the predictive performance of the model.

**Figure 5. Ensemble stage 2 – Uncertainty analysis**

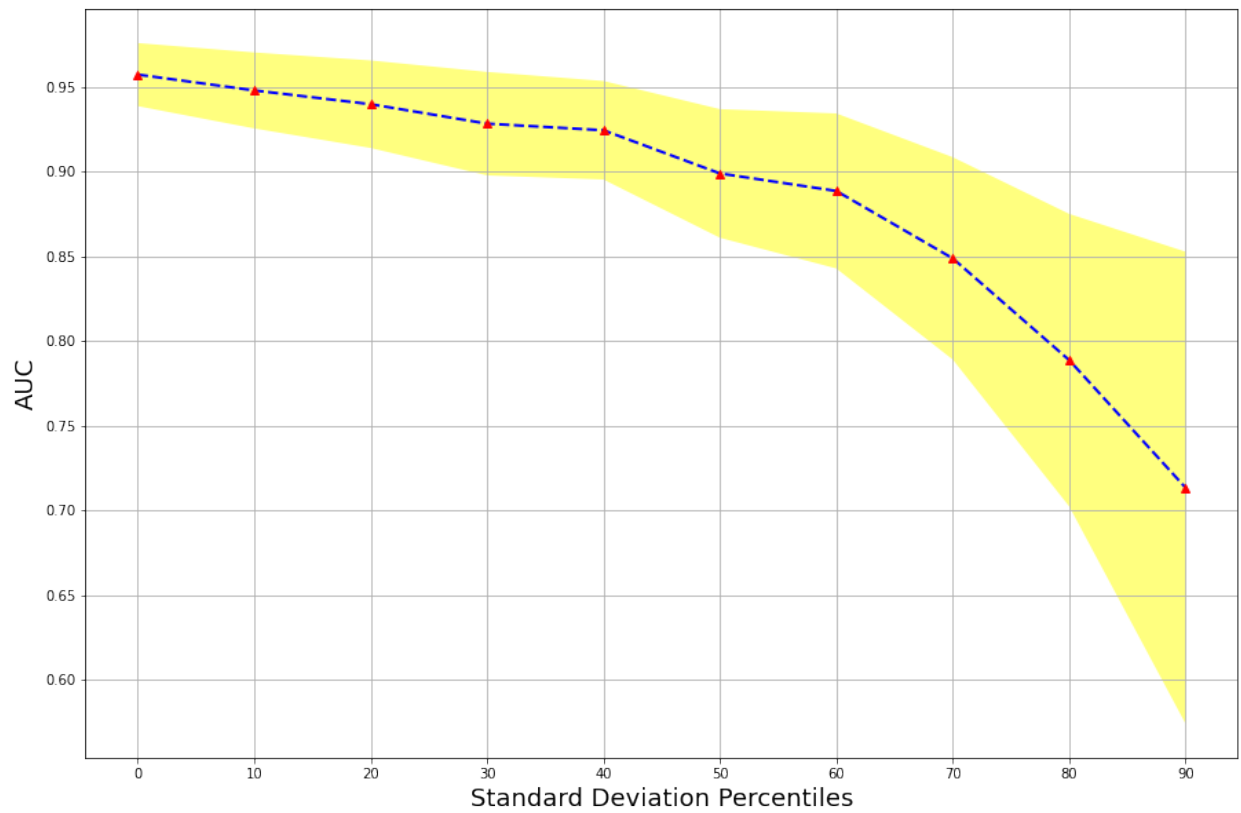

Figure 5 illustrates Ensemble 2's AUC with 95% DeLong Confidence Intervals across standard deviation percentiles or more extreme values. This analysis offers insights into the model's performance variability across varying levels of uncertainty.

**Table 1. Performance metrics of models**

|  | AUC | ACCURACY | SENSITIVITY | SPECIFICITY |
|---|---|---|---|---|
| **ENSEMBLE MODEL 1** | | | | |
| AVG | 0.55488674 | 0.72393939 | 0.19555556 | 0.83429952 |
| STD | 0.13676974 | 0.06450504 | 0.14859008 | 0.05110919 |
| **ENSEMBLE MODEL 2** | | | | |
| AVG | 0.95411165 | 0.91952862 | 0.80777778 | 0.942657 |
| STD | 0.03582502 | 0.0274108 | 0.13361544 | 0.02240635 |
| **STAGED MODEL** | | | | |
| AVG | 0.848595 | 0.861077 | 0.557778 | 0.924879 |
| STD | 0.091797 | 0.046373 | 0.233821 | 0.039411 |

Table 1 summarizes the AVG performance metrics, such as AUC, accuracy, sensitivity, and specificity, for the various models. It includes STD values to indicate metric variability across evaluations. **AVG**: Average, **STD**: Standard Deviation

Table 2. Baseline statistics of categorical features related to hip Fracture

| FEATURE | HIP FRACTURE (NO) | HIP FRACTURE (YES) |
|---|---|---|
| ALCOHOL CONSUMPTION (NEVER CONSUMED) | 1 (0.2%) | 4 (4.3%) |
| ALCOHOL CONSUMPTION (PREVIOUS) | 3 (0.7%) | 0 (0%) |
| ALCOHOL CONSUMPTION (CURRENT) | 449 (99.1%) | 90 (95.7%) |
| AVERAGE HOUSEHOLD INCOME(DO NOT KNOW) | 11 (2.4%) | 2 (2.1%) |
| AVERAGE HOUSEHOLD INCOME (PREFER NOT TO ANSWER) | 24 (5.3%) | 7 (7.4%) |
| AVERAGE HOUSEHOLD INCOME (<18,000£) | 63 (13.9%) | 23 (24.5%) |
| AVERAGE HOUSEHOLD INCOME (18,000£ TO 30,999£) | 130 (28.7%) | 22 (23.4%) |
| AVERAGE HOUSEHOLD INCOME (31,000£ TO 51,999£) | 130 (28.7%) | 20 (21.3%) |
| AVERAGE HOUSEHOLD INCOME (52,000£ TO 100,000£) | 76 (16.8%) | 17 (18.1%) |
| AVERAGE HOUSEHOLD INCOME (>100,000£) | 19 (4.2%) | 3 (3.2%) |
| VARIATION IN DIET (NEVER/RARELY) | 153 (33.8%) | 34 (36.2%) |
| VARIATION IN DIET (SOMETIMES) | 258 (57.0%) | 51 (54.3%) |
| VARIATION IN DIET (OFTEN) | 42 (9.3%) | 9 (9.6%) |
| FALLS IN LAST YEAR (PREFER NOT TO ANSWER) | 1 (0.2%) | 1 (1.1%) |
| FALLS IN LAST YEAR (NONE) | 356 (78.6%) | 74 (78.7%) |
| FALLS IN LAST YEAR (ONLY ONE) | 65 (14.3%) | 13 (13.8%) |
| FALLS IN LAST YEAR (MORE THAN ONE) | 31 (6.8%) | 6 (6.4%) |
| FRACTURE/BROKEN BONES IN LAST 5 YEARS (DO NOT KNOW) | 1 (0.2%) | 0 (0%) |
| FRACTURE/BROKEN BONES IN LAST 5 YEARS (PREFER TO KNOW) | 1 (0.2%) | 0 (0%) |
| FRACTURE/BROKEN BONES IN LAST 5 YEARS (NONE) | 420 (92.7%) | 86 (91.5%) |
| FRACTURE/BROKEN BONES IN LAST 5 YEARS (YES) | 31 (6.8%) | 8 (8.5%) |
| GENETIC SEX (FEMALE) | 227 (50.1%) | 54 (57.4%) |
| GENETIC SEX (MALE) | 226 (49.9%) | 40 (42.6%) |
| MAJOR CHANGE IN DIET IN LAST 5 YEARS(NOT DUE TO ILLNESS) | 283 (62.5%) | 65 (69.1%) |
| MAJOR CHANGE IN DIET IN LAST 5 YEARS (DUE TO ILLNESS) | 29 (6.4%) | 7 (7.4%) |
| MAJOR CHANGE IN DIET IN LAST 5 YEARS (OTHER REASONS) | 141 (31.1%) | 22 (23.4%) |
| SEX (FEMALE) | 227 (50.1%) | 54 (57.4%) |
| SEX (MALE) | 226 (49.9%) | 40 (42.6%) |
| SMOKING (PREFER NOT TO ANSWER) | 3 (0.7%) | 0 (0%) |
| SMOKING (NEVER) | 247 (54.5%) | 52 (55.3%) |
| SMOKING (PREVIOUS) | 187 (41.3%) | 36 (38.3%) |
| SMOKING (CURRENT SMOKER) | 16 (3.5%) | 6 (6.4%) |
| VITAMIN SUPPLEMENT (NONE) | 377 (83.2%) | 79 (84.0%) |
| VITAMIN SUPPLEMENT (YES) | 76 (16.8%) | 15 (16.0%) |

Table 2 displays the information on different factors related to hip fractures, comparing individuals who experienced hip fractures ("Hip Fracture (Yes)") with those who didn't ("Hip Fracture (No)"). Each row represents a specific feature mentioned in the study. The numbers in the table represent percentages and counts within each group.

**Table 3. Baseline statistics of continuous features related to hip fracture**

| FEATURE | MEAN | | STANDARD DEVIATION | |
|---|---|---|---|---|
| | HIP FRACTURE (NO) | HIP FRACTURE (YES) | HIP FRACTURE (NO) | HIP FRACTURE (YES) |
| AGE | 57.70 | 59.70 | 7.05 | 7.51 |
| FEMUR NECK BMC(LEFT) | 4.96 | 4.48 | 0.99 | 1.12 |
| FEMUR NECK BMD(LEFT) | 0.94 | 0.82 | 0.14 | 0.13 |
| FEMUR TOTAL BMD(LEFT) | 1.01 | 0.86 | 0.16 | 0.13 |
| FEMUR TOTAL BMD T-SCORE(LEFT) | -0.32 | -1.47 | 1.15 | 0.92 |
| FEMUR TROCH BMD(LEFT) | 0.85 | 0.72 | 0.16 | 0.14 |
| FEMUR TROCH BMD T-SCORE(LEFT) | -0.10 | -1.21 | 1.25 | 1.06 |
| FEMUR WARDS BMD (LEFT) | 0.73 | 0.60 | 0.14 | 0.12 |
| FEMUR WARDS BMD T-SCORE(LEFT) | -1.59 | -2.51 | 1.08 | 0.89 |
| PELVIS BMC | 334.96 | 299.12 | 79.90 | 73.75 |
| FEMUR NECK BMC(RIGHT) | 4.99 | 4.43 | 1.00 | 0.89 |
| FEMUR NECK BMD(RIGHT) | 0.94 | 0.82 | 0.14 | 0.11 |
| FEMUR TOTAL BMD(RIGHT) | 1.00 | 0.86 | 0.16 | 0.13 |
| FEMUR NECK BMD T-SCORE(RIGHT) | -0.34 | -1.43 | 1.14 | 0.94 |
| FEMUR TROCH BMD(RIGHT) | 0.84 | 0.72 | 0.16 | 0.14 |
| FEMUR TROCH BMD T-SCORE(RIGHT) | -0.15 | -1.22 | 1.23 | 1.09 |
| FEMUR WARDS BMD (RIGHT) | 0.73 | 0.60 | 0.15 | 0.15 |
| FEMUR WARDS BMD T-SCORE(RIGHT) | -1.58 | -2.55 | 1.13 | 0.85 |
| WEIGHT | 77.27 | 74.48 | 14.39 | 15.05 |

Table 3 compares measurements and statistics between two groups: individuals who experienced hip fractures and those who did not. Each row represents a specific feature. The columns show the average value (**mean**) and the variation (**standard deviation**) within each group.

**TABLE 4. Categorical features p-values from chi square test and fisher test**

| FEATURE | CHI-SQUARE | FISHER'S |
|---|---|---|
| FALLS IN LAST YEAR (PREFER NOT TO ANSWER) | 0.7691 | 0.3144 |
| FALLS IN LAST YEAR (NONE) | 0.9999 | 1.0000 |
| FALLS IN LAST YEAR (ONLY ONE) | 0.9999 | 1.0000 |
| FALLS IN LAST YEAR (MORE THAN ONE) | 0.9999 | 1.0000 |
| MAJOR CHANGE IN DIET IN LAST 5 YEARS (NOT DUE TO ILLNESS) | 0.2684 | 0.2402 |
| MAJOR CHANGE IN DIET (DUE TO ILLNESS) | 0.8860 | 0.6522 |
| MAJOR CHANGE IN DIET (OTHER REASONS) | 0.1721 | 0.1725 |
| VARIATION IN DIET (NEVER/RARELY) | 0.7444 | 0.7201 |
| VARIATION IN DIET (SOMETIMES) | 0.7144 | 0.6488 |
| VARIATION IN DIET (OFTEN) | 0.9999 | 0.8480 |
| ALCOHOL CONSUMPTION (NEVER CONSUMED) | 0.0017 | 0.0036 |
| ALCOHOL CONSUMPTION (PREVIOUS) | 0.9810 | 1 |
| ALCOHOL CONSUMPTION (CURRENT) | 0.0448 | 0.0329 |
| SMOKING (PREFER NOT TO ANSWER) | 0.9810 | 1 |
| SMOKING (NEVER) | 0.9786 | 0.9098 |
| SMOKING (PREVIOUS) | 0.6744 | 0.6452 |
| SMOKING (CURRENT SMOKER) | 0.3213 | 0.2428 |
| AVERAGE HOUSEHOLD INCOME (DO NOT KNOW) | 0.9999 | 1.0000 |
| AVERAGE HOUSEHOLD INCOME (PREFER NOT TO ANSWER) | 0.5654 | 0.4599 |
| AVERAGE HOUSEHOLD INCOME (<£18,000) | 0.0162 | 0.0185 |
| AVERAGE HOUSEHOLD INCOME (£18,000 TO £30,999) | 0.3596 | 0.3148 |
| AVERAGE HOUSEHOLD INCOME (£31,000 TO £51,999) | 0.1800 | 0.1628 |
| AVERAGE HOUSEHOLD INCOME (£52,000 TO £100,000) | 0.8757 | 0.7635 |
| AVERAGE HOUSEHOLD INCOME (>£100,000) | 0.8714 | 1.0000 |
| VITAMIN SUPPLEMENT | 0.9665 | 1.0000 |
| GENETIC SEX | 0.2373 | 0.2132 |
| FRACTURE/BROKEN BONES IN LAST 5 YEARS | 0.8658 | 0.6689 |
| SEX | 0.2373 | 0.2132 |

Table 4 presents p-values from both the chi-square test and Fisher test for categorical features. The tests evaluate associations between categorical variables and the outcome.

**TABLE 5. Categorical features p-values from t-test**

| FEATURE | P-VALUE |
|---|---|
| AGE | 0.0188 |
| WEIGHT | 0.1012 |
| RIGHT FEMUR NECK BMD | <0.0001 |
| RIGHT FEMUR NECK BMC | <0.0001 |
| RIGHT FEMUR TOTAL BMD | <0.0001 |
| RIGHT FEMUR TOTAL BMD T-SCORE | <0.0001 |
| RIGHT TROCHANTER BMD | <0.0001 |
| RIGHT TROCHANTER BMD T-SCORE | <0.0001 |
| RIGHT WARDS BMD | <0.0001 |
| RIGHT WARDS BMD T-SCORE | <0.0001 |
| LEFT FEMUR NECK BMD | <0.0001 |
| LEFT FEMUR NECK BMC | 0.00019 |
| LEFT FEMUR TOTAL BMD | <0.0001 |
| LEFT FEMUR TOTAL BMD T-SCORE | <0.0001 |
| LEFT TROCHANTER BMD | <0.0001 |
| LEFT TROCHANTER BMD T-SCORE | <0.0001 |
| LEFT WARDS BMD | <0.0001 |
| LEFT WARDS BMD T-SCORE | <0.0001 |
| PELVIS BMC | <0.0001 |

Table 5 displays the p-values obtained from t-tests for categorical features used in the study. The t-tests evaluate differences in means between groups for each variable.

# Supplementary Materials

## Table 1. List of clinical variables used in this study.

| Feature | Meaning | Range |
|---|---|---|
| **Age** | Age of the participant on the day of attending an Initial Assessment Centre, truncated to whole year. | Integer, years |
| **Sex** | Gender of the participant. | Categorical Female (0), Male (1) |
| **Genetic sex** | Biological sex of the participant. | Categorical Female (0), Male (1) |
| **Weight** | Weight measured during the initial assessment center visit and amalgamated into a single item. | Continuous, Kg |
| **Average total household income before tax** | Total household income before tax received by the household, categorized into income brackets. | Categorical (Various income ranges) |
| **Smoking status** | Current or past smoking status of the participant. | Categorical (Never, Previous, Current) |
| **Alcohol drinker status** | Current or past alcohol consumption status of the participant. | Categorical (Never, Previous, Current) |
| **Variation in diet** | Frequency of diet variation. | Categorical (Never/rarely, Sometimes, Often) |
| **Major dietary changes in the last 5 years** | Major changes made to the diet in the last 5 years. | Categorical (No, Yes because of illness, Yes because of other reasons) |
| **Falls in the last year** | Frequency of falls in the last year. | Categorical (No falls, Only one fall, More than one fall) |
| **Fractured/broken bones in last 5 years** | History of fractured or broken bones in the last 5 years. | Categorical: Yes, No, Do not know, Prefer not to answer |
| **Femur neck BMD (right)** | Bone mineral density of the right femur neck. | Continuous, g/cm2 |
| **Femur neck BMC (right)** | Bone mineral content of the right femur neck. | Continuous, g |
| **Femur total BMD (right)** | Bone mineral density of the right femur. | Continuous, g/cm2 |
| **Femur total BMD T-score (right)** | T-score of the right femur total bone mineral density. | Continuous, Std.Devs |
| **Femur troch BMD (right)** | Bone mineral density of the right femur trochanter. | Continuous, g/cm2 |
| **Femur troch BMD T-score (right)** | T-score of the right femur trochanter bone mineral density. | Continuous, Std.Devs |
| **Femur wards BMD (right)** | Bone mineral density of the right femur Ward's triangle. | Continuous, g/cm2 |
| **Femur wards BMD T-score (right)** | T-score of the right femur Ward's triangle bone mineral density. | Continuous, Std.Devs |
| **Femur neck BMD (left)** | Bone mineral density of the left femur neck. | Continuous, g/cm2 |
| **Femur neck BMC (left)** | Bone mineral content of the left femur neck. | Continuous, g |
| **Femur total BMD (left)** | Bone mineral density of the left femur. | Continuous, g/cm2 |
| **Femur total BMD T-score (left)** | T-score of the left femur total bone mineral density. | Continuous, Std.Devs |
| **Femur troch BMD (left)** | Bone mineral density of the left femur trochanter. | Continuous, g/cm2 |
| **Femur troch BMD T-score (left)** | T-score of the left femur trochanter bone mineral density. | Continuous, Std.Devs |
| **Femur wards BMD (left)** | Bone mineral density of the left femur Ward's triangle. | Continuous, g/cm2 |
| **Femur wards BMD T-score (left)** | T-score of the left femur Ward's triangle bone mineral density. | Continuous, Std.Devs |
| **Pelvis BMC** | Bone mineral content of the pelvis. | Continuous, g |
| **Vitamin and mineral supplements** | Regular intake of various vitamins and minerals. | Categorical |